\documentclass[amsmath,amssymb,twocolumn,showpacs]{revtex4}
\usepackage{bm}

\begin{document}

\newcommand{\uu}[1]{\underline{#1}}
\newcommand{\pp}[1]{\phantom{#1}}
\newcommand{\be}{\begin{eqnarray}}
\newcommand{\ee}{\end{eqnarray}}
\newcommand{\ve}{\varepsilon}
\newcommand{\ii}{\iota}
\newcommand{\vs}{\varsigma}
\newcommand{\Tr}{{\,\rm Tr\,}}
\newcommand{\pol}{\textstyle{\frac{1}{2}}}
\newcommand{\lbar}{l_{^{\!\bar{}}}}

\title{Quantum chaotic systems with arbitrarily large Ehrenfest times}
\author{Maciej Kuna}
\affiliation{
Wydzia{\l} Fizyki Technicznej i Matematyki Stosowanej\\
Politechnika Gda\'nska, 80-952 Gda\'nsk,
Poland}

\begin{abstract}
A class of time independent and physically meaningful Hamiltonians leads to evolution of observable quantities whose Ehrenfest times are arbitrarily large. This fact contradicts the popular claim that the true chaos is in quantum mechanics excluded by first principles.
\end{abstract}
\maketitle

In his introductory remarks to one of the conferences on quantum chaos Michael Berry formulated a kind of credo of theorists working on chaotic aspects of quantum systems: ``There is no chaos in quantum mechanics.(...) In all except some very special cases (e.g. the `quantum' system got by regarding the Liouville equation of a chaotic classical system as a Schr\"odinger equation, whose specialness is that its `Hamiltonian' is linear in the `momenta') $\hbar$ smoothes away the fine classical phase-space structure, and prevents chaos from developing. The inaccurate phrase `quantum chaos' is simply shorthand, denoting quantum phenomena characteristic of classically chaotic systems, quantal `reflections' or `parallels' of chaos..." \cite{Berry}. The true chaos --- involving hypersensitivity to initial conditions, strange attractors and the like --- is believed to be excluded from quantum mechanics by first principles. The orthodox faith of quantum physicists allows for chaos only in the very limited sense of a property of semiclassical approximations. This is why one of the classic textbooks on the subject has only signatures of chaos in the title \cite{Haake}, and an appropriate entry of the current PACS scheme reads: ``Semiclassical chaos (`quantum chaos')".

There are various reasons why chaos is claimed to be impossible in quantum mechanics, but two of them seem most suggestive. First, the dynamics of quantum states is unitary hence linear, while it is known that chaos in autonomous systems occurs for nonlinear evolutions. Secondly, the initial conditions in phase space are not given exactly due to the uncertainty principle for $\bm p$ and $\bm q$; in consequence the ``Ehrenfest time", determining for how long a quantum system can follow a classical chaotic trajectory, cannot be larger than a certain value determined by the Planck constant. The artificial example of a chaotic `quantum' system mentioned by Berry and discussed for the first time in \cite{CIS}, reduces essentially to the following calculation
\be
i\hbar\dot \psi\big(\bm x(t)\big)=\dot{\bm x}(t)\cdot i\hbar\bm \nabla \psi(\bm x)\big|_{\bm x=\bm x(t)}=H\psi\big(\bm x(t)\big)\label{K}
\ee
where $\bm x(t)$ is a solution of a classical problem. The formula has physically nothing to do with quantization although formally it has a Schr\"odinger form with $H$ linear in $\bm p=-i\hbar\bm \nabla$, and the Hamiltonian can be made self-adjoint if one appropriately defines a scalar product.

The goal of this note is to show that, in spite of the above arguments, there exists a class of quantum systems with physically meaningful time-independent Hamiltonians, whose Ehrenfest times can be arbitrarily large. The necessary condition for chaos is here the same as in classical physics: The evolution equation for some {\it observables\/} must be nonlinear.

Indeed, let us consider the time-independent Hamiltonian
\be
H=\frac{\bm P^2}{2M}-\frac{1}{2}\big(\bm x\cdot \bm F+\bm F\cdot \bm x\big)\label{H}
\ee
where $\bm P$ and $\bm F$ commute. The Heisenberg equation of motion
\be
\dot {\bm P}=\frac{1}{i\hbar}[\bm P, H]=\bm F\label{F}
\ee
shows that $\bm F$ is a force. If $\bm F$ is a constant vector then $H$ describes, in particular, the dynamics of an atom of mass $M$ falling freely in gravitational field. Now, we know that the free fall is an idealization and various velocity dependent friction forces often occur. This is true also in the quantum case if, for example, the atom falls in the presence of a laser light. The Hamiltonian (\ref{H}) with $\bm F$ depending on $\bm P$ is the simplest toy model of atomic cooling by light forces.

Friction forces occurring in realistic systems are nonlinear and thus the Heisenberg equation (\ref{F}) is in general nonlinear as well. For example the simple quadratic force
\be
\bm F
&=&
\left(
\begin{array}{c}
-\sigma P_1+ \sigma P_2\\
\tau P_1  - P_2\\
-\beta P_3
\end{array}
\right)
+
P_1
\left(
\begin{array}{c}
0\\
-P_3\\
P_2
\end{array}
\right)
,
\ee
where $\beta$, $\sigma$, $\tau$, are constant parameters, implies
\be
 \dot P_1 &=& \frac{1}{i\hbar} [P_1,H] = \sigma (P_2 - P_1), \\
 \dot P_2 &=& \frac{1}{i\hbar} [P_2, H] = P_1(\tau - P_3) - P_2,\\
 \dot P_3 &=& \frac{1}{i\hbar}[P_3,H] =  P_1 P_2 -\beta P_3,
\ee
which is nothing else but the chaotic Lorenz system \cite{Lor}. The corresponding $H$ is an invariant of the dynamics, but not of the same type as the Ku\'s invariants occurring for some choices of $\beta$, $\sigma$, and $\tau$ \cite{Kus}.

A solution of the above Lorenz system is an operator whose action on momentum-space wave function is
\be
P_k(t)\psi(\bm p) &=& f_k(t,\bm p)\psi(\bm p),
\ee
where $f_k(t,\bm p)$ is a solution of the classical dynamical problem
\be
 \dot f_1 &=& \sigma (f_2 - f_1), \\
 \dot f_2 &=& f_1(\tau - f_3) - f_2,\\
 \dot f_3 &=& f_1 f_2 -\beta f_3,
\ee
with initial conditions $f_k(0,\bm p)=p_k$; $\psi(\bm p)$ is the wave function at $t=0$.
The average value of the evolving observable thus reads
\be
\langle\psi|P_k(t)|\psi\rangle=\int d^3p\, f_k(t,\bm p)|\psi(\bm p)|^2.\label{av}
\ee
The finite Ehrenfest time is typically claimed to be a consequence of the uncertainty principle in phase space, i.e. the minimal volume $\Delta p_j\Delta q_j\geq \hbar$. Eq.~(\ref{av}) explains why the dynamics of average momentum involves in this example an arbitrarily large Ehrenfest time: There is no uncertainty principle limiting the volume $\Delta p_1\Delta p_2\Delta p_3$.
In the limiting case of an eigenstate of momentum operators, where the wave packet shrinks to the Dirac delta centered at $\bm k$, the average follows the trajectory $f_k(t,\bm k)$, i.e. the Ehrenfest time becomes infinite.

The fact that $\hbar$ cannot lead to any upper bound on the Ehrenfest time trivially follows also from the fact that (\ref{av}) is not in any sense related to the Planck constant; $\hbar$ is absent in both the evolution equation (which is just the Lorenz system) and the initial probability density, which is arbitrary. The form of (\ref{av}) is identical to the classical expression for an average trajectory and thus the standard classical estimates  \cite{Schuster} based on the maximal Lyapunov exponent or Kolmogorov-Sinai entropy are valid and can be employed without any modification.

The Hamiltonian we have used cannot be claimed to be unphysical or exotic. One can only complain that it is too simple to be a realistic description of atoms interacting with external fields but, of course, the same happens in the classical theory of Hamiltonian chaos. The celebrated Henon-Heiles system is just a toy model of many-body gravitational interactions \cite{HH}.

$H$ is certainly not linear in momentum (the case mentioned by Berry), although it is linear in $\bm x$. Hamiltonians containing terms linear in some observable occur in so many applications that one can even claim they are generic in quantum mechanics. Many phenomenological quantum models describing interactions with systems whose structure is too complicated to allow for first-principle modeling (e.g. systems interacting with reservoirs) involve such interaction terms. The fact that $H$ contains a third-order interaction is also not strange: Third-order polynomials are the simplest functions that lead to nonlinear Heisenberg equations, and nonlinearity is certainly a necessary condition for chaos, at least in time-independent systems. Translating (\ref{H}) into quantum optical terms we get a three-mode interaction that includes squeezing and two- and three-photon processes. In time-dependent Hamiltonians one can introduce chaos by an appropriate choice of chaotic maps, such as the Arnold cat map discussed by Weigert \cite{W1,W2,W3}, and then the Heisenberg evolution can be piecewise linear. 

Let me try to put all of this in a wider context. How is it possible that in spite of linearity of quantum mechanics we have systems evolving chaotically? The answer is very simple: Heisenberg-picture equations are typically nonlinear --- this is anyway why one can speak of nonlinear quantum optics. Quantum chaos can exist because operator equations for observables can be chaotic. Although certain tools for investigation of chaotic properties at the level of observables were prepared in the literature a long time ago \cite{LE}, one could not find an example of a chaotic Heisenberg equation. The Hamiltonians (\ref{H}) solve the problem in a trivial way \cite{?}.

I'm indebted to Marek Czachor, Adam Majewski, and Stefan Weigert for discussions that helped to improve the argument and its presentation.


\begin{thebibliography}{99}
\bibitem{LE}The first definition of a Lyapunov exponent at the level of Heisenberg-picture equations was discussed in  \cite{MK93} (for further developments cf. \cite{ENTS}; the relations between \cite{MK93} and \cite{ENTS} were discussed in detail in \cite{MM}), but at that time one could not find a convincing example of a chaotic Heisenberg equation, although systems with instabilities were known \cite{MK95}.
\end{thebibliography}
\end{document}